\documentclass[12pt]{article}
\usepackage[margin=1in]{geometry}
\usepackage[utf8]{inputenc}

\usepackage{amsmath,amssymb,amsfonts,amsthm}

\usepackage{floatrow}

\usepackage{breqn}

\usepackage{bold-extra}

\usepackage{graphicx}
\usepackage{xcolor}
\usepackage{url}
\usepackage{hyperref}
\hypersetup{
  colorlinks   = true, 
  urlcolor     = blue, 
  linkcolor    = black, 
  citecolor   = black 
}
\usepackage[numbers,sort&compress]{natbib}
\let\OLDthebibliography\thebibliography
\renewcommand\thebibliography[1]{
  \OLDthebibliography{#1}
  \setlength{\parskip}{4pt}
  \setlength{\itemsep}{0pt plus 0.3ex}
}

\allowdisplaybreaks
\usepackage{tabularx}
\usepackage{bbm}
\usepackage{listings}
\usepackage{longtable}
\usepackage{array}

\usepackage{multicol}
\usepackage{enumitem}

\newcommand\refeq[1]{Eq.~(\ref{#1})}

\newcommand\refta[1]{Tab.~\ref{#1}}
\newcommand\refse[1]{Sect.~\ref{#1}}

\newcommand\citere[1]{Ref.~\cite{#1}}

\newcommand\refap[1]{App.~\ref{#1}}
\def\reffi#1{\mbox{Fig.~\ref{#1}}}

\newcommand{\htb}[1]{{\color{black} #1}}

\newcommand{\htm}[1]{{\color{black} #1}}

\newcommand{\htk}[1]{{\color{black} #1}}

\usepackage{soul}

\newcommand{\gev}{\ \mathrm{GeV}}
\newcommand{\tev}{\ \mathrm{TeV}}

\newcommand{\ii}{\text{i}}

\graphicspath{{Plots/}}

\begin{document}

\thispagestyle{empty}

\def\thefootnote{\fnsymbol{footnote}}

\begin{flushright}
  DESY-22-083
\end{flushright}

\vspace*{1cm}

\begin{center}

  {\Large Direct detection of
  pseudo-Nambu-Goldstone dark matter\\[0.4em]
  in a two Higgs doublet plus singlet
  extension of the SM}

  \vspace{1cm}

Thomas
  Biek\"otter$^1$\footnote{thomas.biekoetter@desy.de},
Pedro
  Gabriel$^{2}$\footnote{ptgabriel@gmail.com},
Mar\'ia Olalla Olea
  Romacho$^1$\footnote{maria.olalla.olea.romacho@desy.de}
and
Rui
  Santos$^{2,3}$\footnote{rasantos@fc.ul.pt}
  
\vspace*{1em}
  
{\textit{
  $^1$Deutsches Elektronen-Synchrotron DESY,
   Notkestr.~85, 22607 Hamburg, Germany\\[0.8em]
  $^2$Centro de F\'isica T\'eorica e Computacional,
  Faculdade de C\^iencias,\\
  Universidade de Lisboa, Campo Grande,
  Edif\'icio C8 1749-016 Lisboa, Portugal\\[0.8em]
  $^3$ISEL -- Instituto Superior de Engenharia de Lisboa,\\
  Instituto Polit\'ecnico de Lisboa,
  1959-007 Lisboa, Portugal
}}

\vspace*{1cm}

\begin{abstract}
We calculate the leading radiative corrections
to the dark-matter-nucleon
scattering in the pseudo-Nambu-Goldstone dark
matter model augmented
with a second Higgs doublet (S2HDM).
In this model, the cross sections
\htm{for the scattering of the
dark-matter on nuclei}
vanishes at tree-level in the limit of
zero momentum-transfer due to a
U(1) symmetry. However, this symmetry is softly
broken in order to give a mass to the dark-matter
particle. As a consequence,
non-vanishing scattering cross sections
arise at the loop level.
We find that the current
\htm{cross-section limits from dark-matter}
direct-detection \htm{experiments}
can hardly constrain
the parameter space of the S2HDM.
However, the loop-corrected predictions for the
scattering cross sections can be well within the
reach of future direct-detection
experiments.
As a consequence,
future phenomenological
analyses of the S2HDM should take into account
cross-section predictions beyond tree-level and
the experimental constraints from
dark-matter direct-detection experiments.
\end{abstract}

\end{center}

\renewcommand{\thefootnote}{\arabic{footnote}}
\setcounter{footnote}{0} 

\newpage

\section{Introduction}
\label{secintro}

Dark matter (DM) is still today one of the greatest mysteries in physics. Although the first hints of its existence were reported almost 100 years ago, and significant
pieces of evidence have been gathered from different sources, we are nevertheless ignorant of its nature (see~\htm{\citere{Bertone:2016nfn}} for a review). Perhaps the simplest way to be in tune with all experimental results 
is to consider DM as a particle yet to be discovered. There are many ongoing experiments that can provide further directions  in the search for
the correct description of the DM field. However, in order to unmistakably observe a DM candidate, one needs direct detection (DD) experiments that probe the mass and coupling\htm{s}
of the DM particle \htm{with the
Standard Model (SM) particles}
via its interactions with known objects such as nuclei.  As a DM particle interacts with nuclei, light and electric charge are 
\htm{emitted,} 
providing information about energy
and location of the collision.
If we confine ourselves to the mass region of Weekly Interacting Massive Particles (WIMPs) the most restrictive
and up-to-date constraints were obtained by the PandaX-4T~\cite{PandaX-4T:2021bab}, the XENON1T~\cite{XENON:2018voc} and the
LZ~\cite{LZnew} collaborations.
\htm{With the hope of a future DM
detection,}
the significance of DD experiments in identifying 
the DM candidates points to the need of understanding in great detail the DM-nucleon cross sections in the different proposed models. 
\htm{On the other hand, as long as} 
no DM particle is found
\htm{at DD experiments},
any new proposed model  has to
\htm{comply with the resulting upper limits
on the DM-nucleon cross sections.}

There is an interesting class of models where the SM particle content is extended to include an extra complex singlet field \htb{invariant under a softly broken global $U(1)$ symmetry}. 
 Known as Pseudo-Nambu-Goldstone (pNG) DM models, they have the distinct feature of having a negligible DM direct
 detection cross section at leading order (LO) as first reported in~\citere{Gross:2017dan}. In these models, the tree-level DM-nucleon cross section 
 is proportional to the DM velocity which according to the experimentally gathered evidence is negligible when compared to the speed of light.
Therefore, the first relevant contribution to the cross section comes from the one-loop electroweak corrections to the DM-nucleon cross section.
These were calculated in~\citere{Azevedo:2018exj, Ishiwata:2018sdi, Glaus:2020ihj} for this simplest version of the pNG DM model  and shown to be several orders of magnitude above the tree-level result,
\htm{and they were also shown to be several
orders of magnitude larger than the
finite-momentum-transfer corrections
at tree-level~\cite{Azevedo:2018exj}.}
 In fact, \htm{the one-loop corrected
 scattering cross sections} can be of the order of present \htb{experimental limits} and even more so of future DM detection experiments in some regions of the parameter space.

One can consider many different extensions of the simplest version of the pNG DM model, that is, the complex
singlet \htk{scalar} extension of the SM with a softly broken $U(1)$ symmetry. The hallmark of these models
is a negligible tree-level DD cross section but it requires specific patterns of $U(1)$ symmetry breaking. In fact, as shown in~\citere{Azevedo:2018oxv},
for the simple complex singlet extension, soft $U(1)$ breaking terms other than the quadratic ones may spoil the proportionality of the
coupling\htm{s between the pNG DM and the
Higgs states} to the squared Higgs mass\htm{es}, which in turn precludes the tree-level cancellation of the DM-nucleon cross section. However, it is straightforward to add an arbitrary number of doublets and still 
have a negligible tree-level cross section, provided the right pattern of soft symmetry breaking is implemented. Other, more sophisticated extensions with the exact same feature were also discussed in the literature~\cite{Karamitros:2019ewv, Cai:2021evx}.

In this work we consider 
\htm{an} extension of the
\htm{simplest version of the}
pNG DM model 
\htm{containing a second} Higgs doublet\htb{, which entails numerous advantages over the version with only one Higgs doublet. For instance, the former
has been shown to \htm{be able to}
realize a 
first-order EW phase transition~\cite{Biekotter:2021ovi, Zhang:2021alu}
in contrast to the pNG DM model with only
one Higgs doublet \htk{in which
a first-order phase transition can
be realized only if the scalar potential
contains terms that spoil the
tree-level cancellation of the DM-nucleon
scattering cross
sections~\cite{Kannike:2019wsn,Alanne:2020jwx}}.}
The scalar sector of the model consists of two doublets and one complex \htm{scalar} singlet
and we will refer to this extension as S2HDM - the singlet extension of the of the two-Higgs doublet model. The DM candidate originates from the
\htm{imaginary part of the}
singlet field while the
\htm{real component of the singlet
field aquires a vacuum expectaion
value (vev) and} mixes with the
other two CP-even fields from the two doublets. While the pseudoscalar and charged sectors phenomenology is very similar to that of the 2HDM, there are now three CP-even fields and one DM particle.
The model was already studied in~\citere{Jiang:2019soj,Zhang:2021alu,Biekotter:2021ovi}.
\htm{In \citere{Biekotter:2021ovi}
the S2HDM was confronted with the relevant
theoretical and experimental constraints, and,
the interplay between the
collider phenomenology and the dark-matter
phenomenology was investigated in detail.}

Since \htk{in the S2HDM} the DD cross section
is again negligible at tree-level, the
leading term\htm{s are} given by the one-loop electroweak contribution to the cross section. It was conjectured in~\citere{Jiang:2019soj} 
that current and future DD experiments will not be able to probe the S2HDM.
\htm{However,} our calculation of the
\htm{next-to leading order}
(NLO) electroweak correction to the DM direct detection cross section will show
that a significant portion of the parameter space will be probed in future DD experiments.

In our calculations we follow closely the procedure in~\citere{Azevedo:2018oxv}.
The outline of the paper is as follows. Section~\ref{sec:2hdm} contains a brief description of the model and introduces our
notation. In section~\ref{sec:calcu} we calculate the electroweak corrections to the spin-independent direct detection cross section. In section \ref{sec:num}, the results
are presented and discussed. Finally, we present our conclusions in section~\ref{conclu}.

\section{The S2HDM}
\label{sec:2hdm}
In order to define
our conventions and 
notation, we briefly 
review in this section the S2HDM, a 2HDM extended by a complex gauge singlet field that is charged under
a softly 
broken global U(1) 
symmetry. For further details
on the model and its phenomenology
we refer the reader to \citere{Biekotter:2021ovi}.

\subsection{Model definitions and notation}
\label{sec:modeldefs}
The tree-level scalar potential 
of the two electroweak Higgs 
doublets $\phi_1$ and $\phi_2$ 
and the complex singlet 
field $\phi_S$ is given by
\begin{align}
V &=
\mu_{11}^2 \left( \phi_1^\dagger \phi_1 \right)
+ \mu_{22}^2 \left( \phi_2^\dagger \phi_2 \right)
- \mu_{12}^2 \left( \left( \phi_1^\dagger \phi_2 \right)
    + \left( \phi_2^\dagger \phi_1 \right) \right)
+ \frac{1}{2} \mu_S^2 \left| \phi_S \right|^2
- \frac{1}{4} \mu_\chi^2 \left( \phi_S^2
    + \left(\phi_S^*\right)^2 \right) \notag \\
&+ \frac{1}{2} \lambda_1 \left( \phi_1^\dagger \phi_1 \right)^2
+ \frac{1}{2} \lambda_2 \left( \phi_2^\dagger \phi_2 \right)^2
+ \lambda_3 \left( \phi_1^\dagger \phi_1 \right)
    \left( \phi_2^\dagger \phi_2 \right)
+ \lambda_4 \left( \phi_1^\dagger \phi_2 \right)
    \left( \phi_2^\dagger \phi_1 \right) \notag \\
&+ \frac{1}{2} \lambda_5 \left(
    \left( \phi_1^\dagger \phi_2 \right)^2
        + \left( \phi_2^\dagger \phi_1 \right)^2
\right)
+ \frac{1}{2} \lambda_6 \left( |\phi_S|^2 \right)^2
+ \lambda_7 \left( \phi_1^\dagger \phi_1 \right) | \phi_S |^2
+ \lambda_8 \left( \phi_2^\dagger \phi_2 \right) | \phi_S |^2
\ ,
\label{eqscalpot}
\end{align}
where the absence of
explicit CP violation
was assumed and
all the Lagrangian
parameters are considered
to be real.
Here the terms that
uniquely involve the
doublet fields 
are identical to
the scalar potential
of the 2HDM with a 
softly broken $\mathbb{Z}_2$ 
symmetry that prevents
the existence of flavour-changing 
neutral Higgs currents at 
tree-level. Depending on the assigned $\mathbb{Z}_2$ charges of the fermions, this results in the typical four Yukawa types (see e.g. \citere{Branco:2011iw} for details).
The terms involving the singlet
field respect a global U(1) symmetry,
which is softly broken
by the term proportional
to $\mu_{\chi}^2$,
providing a non-zero mass 
for the pNG DM.
After EW symmetry breaking
the two Higgs doublets and
the singlet field acquire real
VEVs, about which the Higgs fields
can be expanded in terms of the
charged fields $\phi_{1,2}^+$
and the neutral CP-even ($\rho_{1,2,S}$)
and CP-odd fields ($\sigma_{1,2}$ and $\chi$) as
\begin{equation}
\phi_1 =
\begin{pmatrix}
\phi_1^+ \\
\left( v_{1} + \rho_1 + \mathrm{i} \sigma_1 \right)
    / \sqrt{2}
\end{pmatrix} \ , \quad
\phi_2 =
\begin{pmatrix}
\phi_2^+ \\
\left( v_{2} + \rho_2 + \mathrm{i} \sigma_2 \right)
    / \sqrt{2}
\end{pmatrix} \ , \quad
\phi_S =
\left( v_{S} + \rho_S + \mathrm{i} \chi \right)
    / \sqrt{2} \ .
\end{equation}
The doublets vevs define the EW scale
$v=\sqrt{v_{1}^2 + v_{2}^2} \approx 246 \gev$ and $\tan{\beta} = v_{2}/v_{1}$.
Assuming the previous
vacuum configuration,
the three CP-even fields mix and give
rise to the three 
mass eigenstates 
$h_{a,b,c}$. The mixing in the CP-even
sector is described by an orthogonal
transformation $R$
such that,
\begin{equation}
\begin{pmatrix}
h_a \\ h_b \\ h_c
\end{pmatrix} =
R \cdot
\begin{pmatrix}
\rho_1 \\ \rho_2 \\ \rho_S
\end{pmatrix} \ , \text{ with }
R=
\begin{pmatrix}
c_{\alpha_1}c_{\alpha_2} &
  s_{\alpha_1}c_{\alpha_2} &
    s_{\alpha_2} \\
-(c_{\alpha_1}s_{\alpha_2}s_{\alpha_3}+s_{\alpha_1}c_{\alpha_3}) &
  c_{\alpha_1}c_{\alpha_3}-s_{\alpha_1}s_{\alpha_2}s_{\alpha_3}  &
    c_{\alpha_2}s_{\alpha_3} \\
-c_{\alpha_1}s_{\alpha_2}c_{\alpha_3}+s_{\alpha_1}s_{\alpha_3} &
-(c_{\alpha_1}s_{\alpha_3}+s_{\alpha_1}s_{\alpha_2}c_{\alpha_3}) &
c_{\alpha_2}c_{\alpha_3}
\end{pmatrix}~,
\label{mixingmatrix}
\end{equation}
where $- \pi / 2 \leq \alpha_1, \alpha_2, \alpha_3
\leq \pi / 2$ are the three mixing angles, and
we use the short-hand
notation $s_x = \sin x$, $c_x = \cos x$.
In addition to the notation $h_{a,b,c}$ for
the CP-even scalar states, we will also
make use of the notation $h_{1,2,3}$, where
the mass ordering $m_{h_1} \leq m_{h_2} \leq m_{h_3}$
is implied. Hence, the states $h_{1,2,3}$ can
(in principle) correspond to any of
the states $h_{a,b,c}$ whose decomposition
in terms of the gauge eigenstates $\rho_{1,2,S}$
is defined by the mixing angles $\alpha_{1,2,3}$
as shown in \refeq{mixingmatrix}.
The charged scalar sector
is left unchanged
as compared to the 
2HDM, including
two physical charged
Higgs bosons $H^{\pm}$ with mass
$m_{H^{\pm}}$ and two charged Goldstone bosons $G^{\pm}$ that give rise to the 
longitudinal degrees of freedom of the $W^{\pm}$
bosons after EW 
\htm{symmetry breaking}. 
The pseudoscalar 
fields $\sigma_{1,2}$ produce 
\htm{a}
physical mass state $A$ with 
respective mass
$m_{A}$,
and a neutral  Goldstone boson $G^{0}$ that
constitutes the longitudinal 
polarisation of the $Z$ boson.
\htm{The imaginary component $\chi$ of the
singlet field does not mix with the other
scalar and pseudoscalar fields provided it
does not have a vev.}
Note that 
the remaining
\textit{dark CP-symmetry}
$\phi_{S} \rightarrow \phi_{S}^{*}$ prevents $\chi$ 
from decaying, making it a 
suitable dark 
matter candidate.
The previous 
definitions allow
us to replace
the Lagrangian
parameters by a set of more physically
meaningful parameters that we
will use to sample
the parameter space
of the S2HDM,
\begin{equation}
m_{h_{a,b,c}} \ , \quad
m_A \ , \quad
m_{H^\pm} \ , \quad
m_\chi \ , \quad
\alpha_{1,2,3} \ , \quad
\tan\beta \ , \quad
M = \sqrt{\mu_{12}^2 / \left(
    s_\beta c_\beta \right) } \ , \quad
v_S \ .
\label{eqparas}
\end{equation}
The relations between this set
of parameters and 
the Lagrangian parameters can be found in \citere{Biekotter:2021ovi}.

\subsection{Theoretical constraints}
\label{sec:thecon}

We required the 
tree-level 
scalar potential to be
bounded-from-below (BfB)
along all field directions
\htm{by applying}
the boundedness
from below 
conditions that
were found in \citere{Muhlleitner:2016mzt}.
Additionally, we ensured
the EW minimum to be
the global minimum of the
tree-level scalar potential
to prevent the EW vacuum from
decaying into other unphysical minima 
and from being potentially short-lived
as compared to the age of the universe. The algorithm to
verify for each parameter point whether there
exists a global minimum of the
potential with $v_{1},v_{2},v_{S}>0$ and vanishing
charge-breaking and CP-breaking vevs is
described in \citere{Biekotter:2021ovi}.
Finally, we required that the perturbative treatment
of the model for a given parameter point
is viable. Tree-level perturbative unitarity is
achieved by imposing that the eigenvalues of
the $2 \times 2$ scalar scattering matrix in
the high-energy limit are below an absolute upper
value given by $8 \pi$~\cite{Biekotter:2021ovi}. 
These constraints
 \htm{give rise to} 
upper limits on the absolute values of the quartic
couplings \htm{$\lambda_i$} and combinations thereof.
These conditions are especially 
relevant when there are large mass 
splittings between one of the scalars 
$h_{i}$ and the heavy doublet states $A$, $H^{\pm}$.

\subsection{Experimental constraints}
\label{sec:expcon}
In this section we
will briefly describe the
experimental constraints that we apply
in our numerical discussion.
For a more detailed discussion
we refer to \citere{Biekotter:2021ovi}.
Regarding the Higgs sector, we required 
agreement with the
95$\%$ confident level cross-section
limits 
from collider searches for additional
scalar states by using \texttt{HiggsBounds v.~5.10.2}~\cite{Bechtle:2008jh,
Bechtle:2011sb,Bechtle:2013gu,Bechtle:2013wla,
Bechtle:2015pma,Bechtle:2020pkv}.
Furthermore, we checked for the compatibility with the 
signal-rate measurements of the
SM-like Higgs boson $h_{125}$ 
making use of the public code
\texttt{HiggsSignals v.~2.6.2} ~\cite{Bechtle:2013xfa,
Stal:2013hwa,Bechtle:2014ewa,Bechtle:2020uwn}.
In case of DM masses below $125 / 2 \gev$,
the additional 
decay mode $h_{125} \rightarrow \chi \chi$ suppresses the 
ordinary decays of $h_{125}$ into SM final states.
Constraints derived from
the branching ratio of the invisible decay of the
SM-like Higgs boson
were indirectly imposed by
\texttt{HiggsSignals v.~2.6.2}
through the global 
constraints on the measured 
signal rates of the SM-like 
Higgs.
Further constraints 
related to the presence
of \htm{additional} 
scalar states
are derived from measurements
of electroweak precision 
observables (EWPO).
We applied a two-dimensional
$\chi^2$ test to the oblique
parameters $S$ and $T$~\cite{Peskin:1990zt,Peskin:1991sw}
and discarded parameter points 
for which the predicted values 
were not in agreement with the 
experimental fit result~\cite{Haller:2018nnx} 
at the 95$\%$ confidence level.\footnote{The
fit values for $S$ and $T$ do not take into
account the new measurement of the mass of
the $W$ boson by the CDF
collaboration~\cite{CDF:2022hxs},
which is in significant tension
with the SM prediction.
The new CDF measurement, or a future
world average value including all
exisiting measurement of $M_W$, can be
accommodated in the S2HDM if sizable mass
splittings between the BSM scalar states
are present~\cite{Biekotter:2022abc}.}
Constraints from flavour-physics
observables (see, for instance,
\citere{Haller:2018nnx}) were taken into account by
using a lower limit of $\tan\beta \geq 1.5$,
and for the scan in type~II we additionally
used a lower limit on the mass of the
charged Higgs boson of $m_{H^\pm} \geq 650\gev$.

Finally, we applied several constraints
related to the presence of the dark matter candidate $\chi$.
One of the most important requirements
is to ensure a DM relic abundance $\Omega h^2$
prediction in agreement with the observations
made by the Planck satellite, leading to a
measurement of $(\Omega h^2)_{\rm Planck} = (0.119 \pm
0.003)$~\cite{Planck:2018vyg}.
In our analysis, this value will serve as 
an upper limit for the dark matter relic 
abundance produced via the freeze-out
mechanism predicted by our model,
considering that a prediction lying 
below $(\Omega h^2)_{\rm Planck}$ would allow
for other contributions
(particle or astrophysical) to the dark matter
relic abundance. We computed $\Omega h^2$ using
\texttt{MicrOmegas}~\cite{Belanger:2018ccd}.
Regarding DM direct detection
constraints, the primary objective 
of this paper is to 
compute the DM-nucleon 
scattering cross section 
in the zero momentum-transfer 
at the one-loop level of 
perturbative expansion and to
compare our results 
with the current limits
set by XENON1T~\cite{XENON:2018voc},
\htm{PandaX-4T~\cite{PandaX-4T:2021bab}
and LUX-ZEPLIN (LZ)~\cite{LZnew},}
and, \htm{in addition,} with future limits projected
for DARWIN~\cite{DARWIN:2016hyl}.
Note that there are other planned
direct detection experiments such as
SuperCDMS~\cite{SuperCDMS:2022kse}, just to name
\htm{an example}. We have
taken DARWIN as a prototype \htm{for}
future DD experiment.
We \htk{finally note that we}
do not take into account constraints
from the indirect detection of DM, because
these constraints are only relevant in
a narrow mass window of the DM below
$m_\chi \lesssim 100\gev$~\cite{Fermi-LAT:2016uux},
and the application
of the indirect-detection constraints relies
on a Monte-Carlo simulation which is computationally
quite expensive (see \citere{Biekotter:2021ovi} for details).

\section{Calculation of DM-nucleon scattering cross section}
\label{sec:calcu}
As already mentioned \htm{above}, in the S2HDM
the cross sections for the scattering
of the DM on nuclei vanish in the limit
of zero-momentum exchange at tree-level.
Our goal in this paper is to investigate
whether radiative corrections
to the cross section would make the DM state $\chi$
detectable at current or future
DM direct-detection experiments.
In this section we discuss the
calculation of the radiative corrections,
where we include the dominant contributions
at the one-loop level stemming from diagrams
with the scalar states in the loops.
Our procedure is an extension of the 
calculation performed in Ref.~\cite{Azevedo:2018exj}.
In \refse{sec:num} we will then present the
numerical discussion of the loop-corrected
scattering cross sections in order to answer
the question whether the presence of the
pNG DM state $\chi$ is testable at
\htk{DD} experiments.

\subsection{One-loop contributions to Wilson coefficients}
\label{secwilson}

\begin{figure}
\centering
\includegraphics[height=3cm]{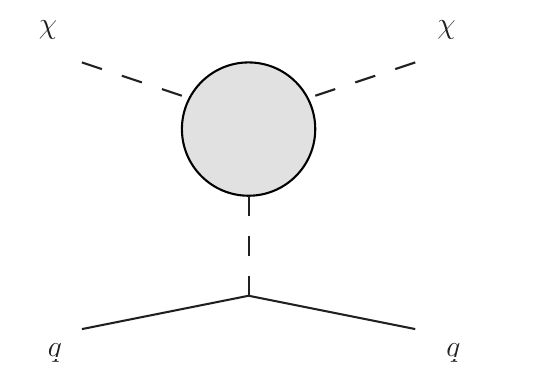}
\includegraphics[height=3cm]{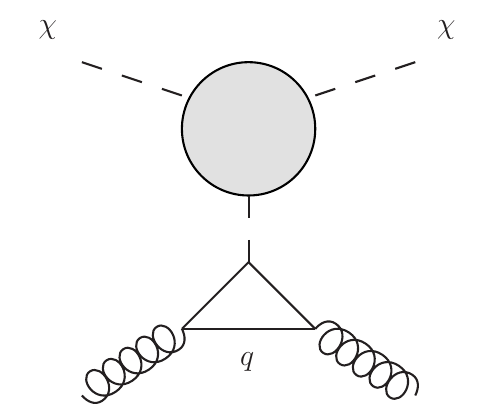}
\includegraphics[height=3cm]{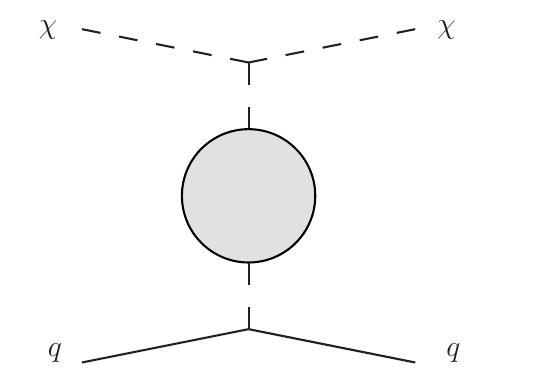}
\includegraphics[height=3cm]{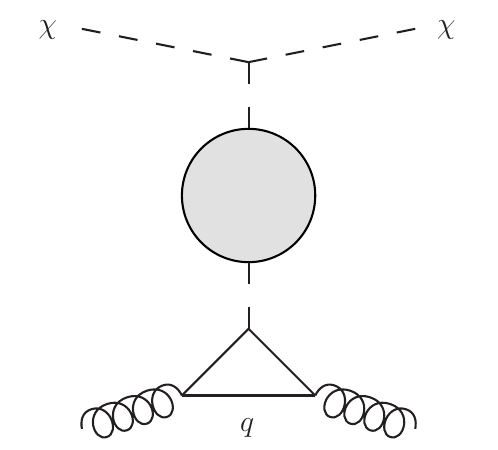}
\caption{Schematic diagrams with upper vertex
corrections and propagator corrections.}
\label{figdiagsII}
\end{figure}

The tree level diagram is just a $t$-channel $\chi q \to \chi q$ scattering where $q$ is a quark belonging to the nucleon. The one-loop
contributions to this process can be divided in three main contributions: upper vertex, lower vertex and mediator corrections. There are also box corrections that do
not fit in this classification. Finally, although of higher order, the gluon initiated processes play a major role in the calculation. The one-loop contributions
considered are the ones given by the topologies schematically shown in \reffi{figdiagsII}. These include only upper vertex and 
mediator corrections. Let us now discuss in detail why the remaining contributions were discarded. 

\begin{figure}
\centering
\includegraphics[width=0.85\textwidth]{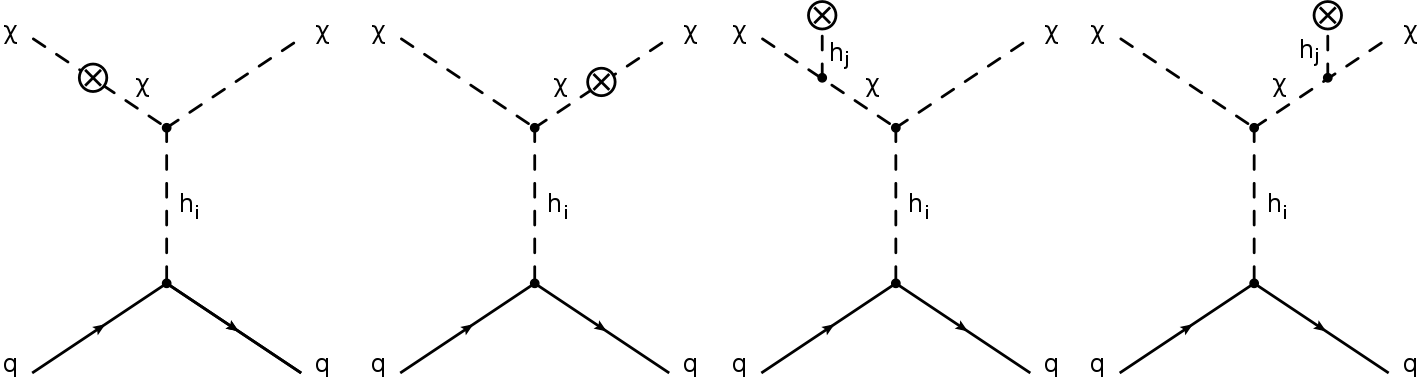}

\vspace{0.5cm}
\includegraphics[width=0.85\textwidth]{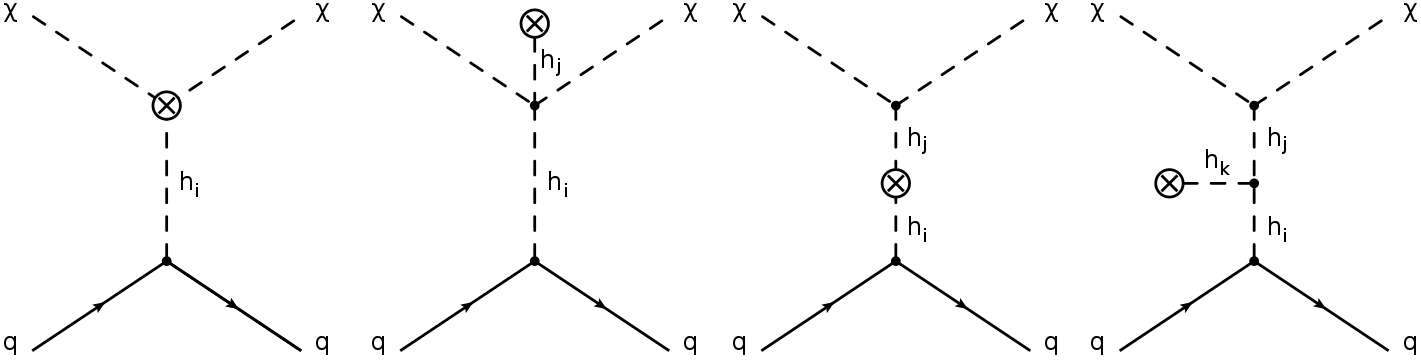}
\caption{Counter-term insertion diagrams for the DM-nucleon scattering ($i,j\in\{1,2,3\}$).}
 \label{fig:counter}
\end{figure}

The tree-level $\chi q \to \chi q$ amplitude vanishes in the limit of zero momentum transfer
\htm{(the explicit expression is given
in \refap{app:treeamp}).}
Hence, the one-loop amplitude has to be finite in the same limit, that is, there is no need for a renormalization
prescription nor for any counterterm. This was already proven in~\citere{Azevedo:2018exj} for the singlet extension and again checked for our model.
We explicitly verified the cancellation of the counterterm diagrams which can be carried out without specifying the individual counterterms, and thus in a generic fashion that
is valid for all four Yukawa types of the S2HDM (counterterms insertions are shown in~\reffi{fig:counter}). As a consequence, the sum of all amplitudes is UV-finite (without the addition of counterterm
diagrams), and the sum is also independent of the renormalization scale, which  we verified numerically.
Our analysis of the UV-finiteness of the
one-loop amplitude is specific to the
S2HDM, although we expect the same result
to hold in a broad class of models
which feature a vanishing tree-level
amplitude in the limit of zero momentum
transfer, because then
there is no counterterm that could cancel a
UV-divergent contribution at one-loop level.

\begin{figure}
\centering
\includegraphics[width=0.85\textwidth]{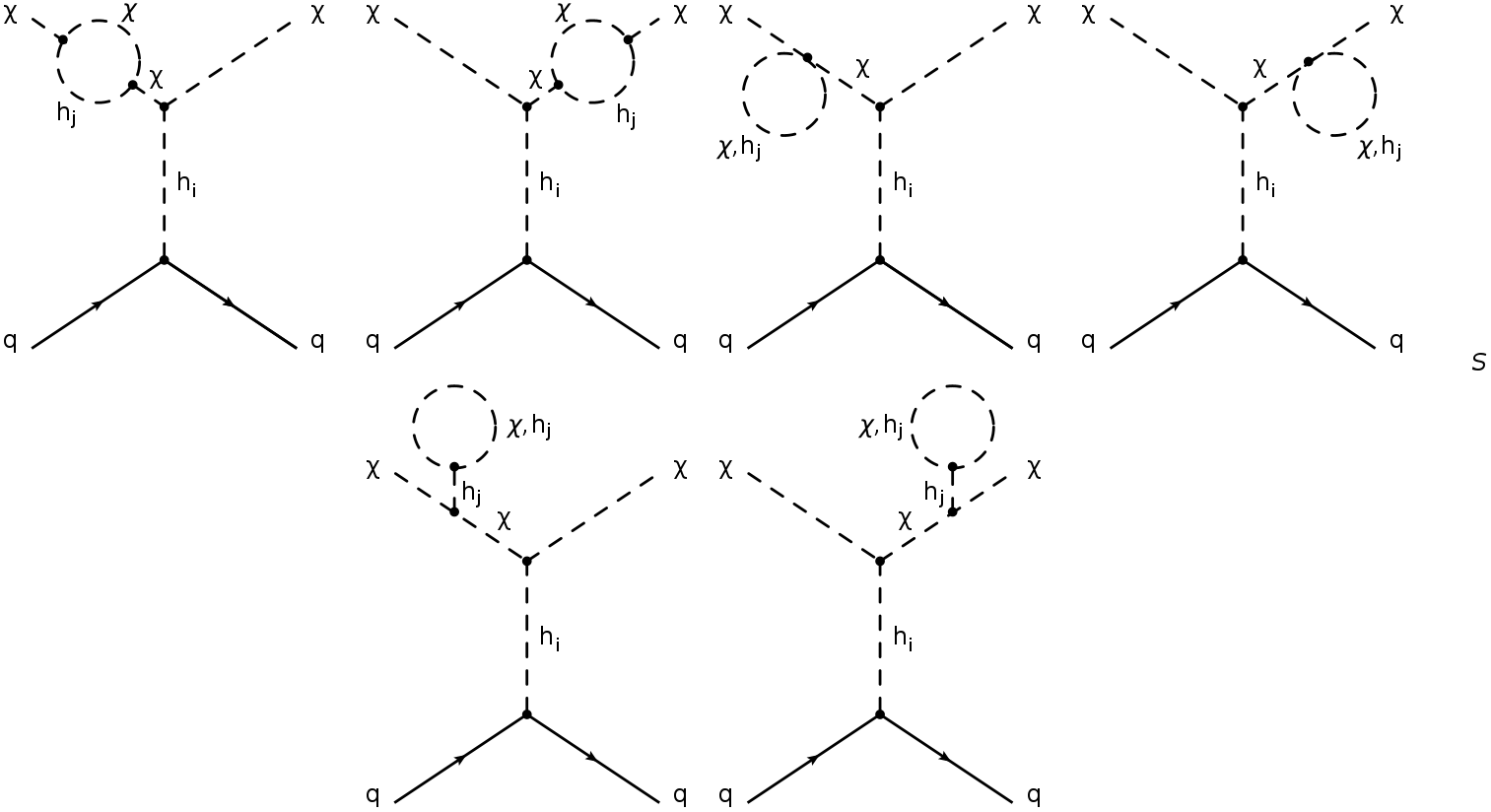}
\caption{External leg corrections to the DM-nucleon scattering with $i,j,k\in\{1,2,3\}$}
\label{fig:exter}
\end{figure}

In~\reffi{fig:exter} we show the corrections on the external
$\chi$-legs.  These corrections vanish in the limit of zero-momentum
transfer, since the corresponding amplitudes are proportional to the tree-level amplitude
which themselves vanish by means of the U(1) symmetry, such that the corresponding
diagrams do not have to be considered.
\begin{figure}
\centering
\includegraphics[width=0.85\textwidth]{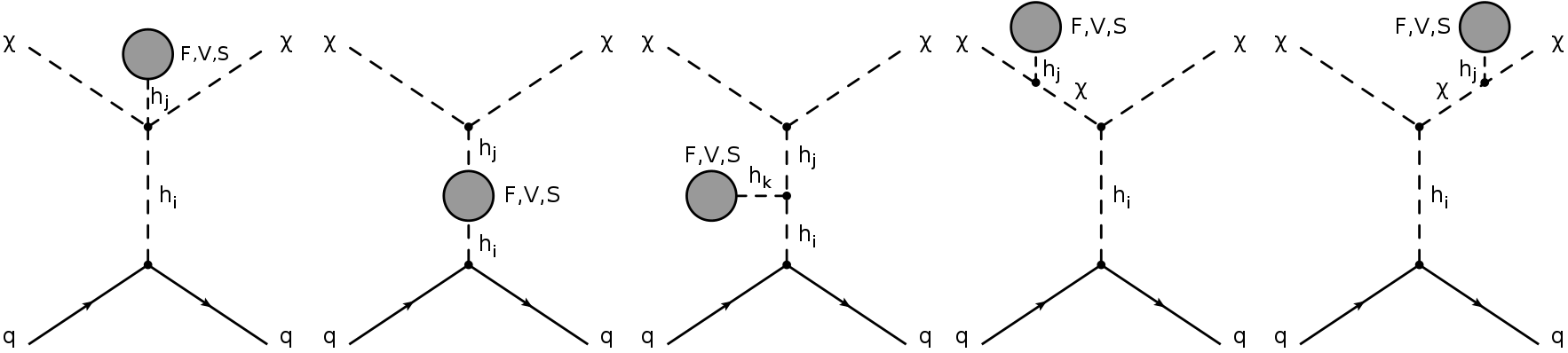}
\caption{One-loop diagrams with loops containing SM particles, $A$ or $H^\pm$ with $F\in\{u,c,s,c,b,t\}$, $V\in\{Z,W\}$ and $S\in\{G_0,G_\pm,A,H^\pm\}$}
\label{fig:SM}
\end{figure}
Finally we present in~\reffi{fig:SM} the set of diagrams with all SM particles, the charged scalars and the pseudoscalar in self-energies and tadpole loops. 
We explicitly verified that only diagrams with the neutral CP-even Higgs bosons and the DM state $\chi$ in the loops give rise to non-zero contributions,
whereas the diagrams with the fermions, the gauge bosons, the pseudoscalar, the charged Higgs bosons and their corresponding Goldstone bosons
in the loop cancel due to the proportionality to the tree-level amplitude. We note that again this was also shown to be true for the complex scalar extension~\cite{Azevedo:2018exj},
but in the S2HDM 
\htm{there are}
new particles in the scalar sector and the proportionality to the tree-level amplitude \htm{is} 
not obvious. 
As a consequence of this result, the one-loop corrections
to the scattering cross section
considered in our analysis are
independent of the gauge fixing,
which we also explicitly verified by
calculating the amplitudes in the
$R_\xi$-gauge and varying the gauge-fixing
parameter.

\htm{A} set of diagrams that we did not take into account are the box contributions
for the process $\chi q \to \chi q$. This is not because the amplitudes are proportional
to the tree-level amplitude but rather because their contribution is at least one order of magnitude 
smaller than the vertex and  mediator contributions. This was checked for two different models~\cite{Glaus:2019itb, Glaus:2020ape, Glaus:2020ihj}
and is mainly related to the fact that the amplitude is proportional to product of two Yukawa couplings to
light quarks.

\begin{figure}
\centering
\includegraphics[height=6cm]{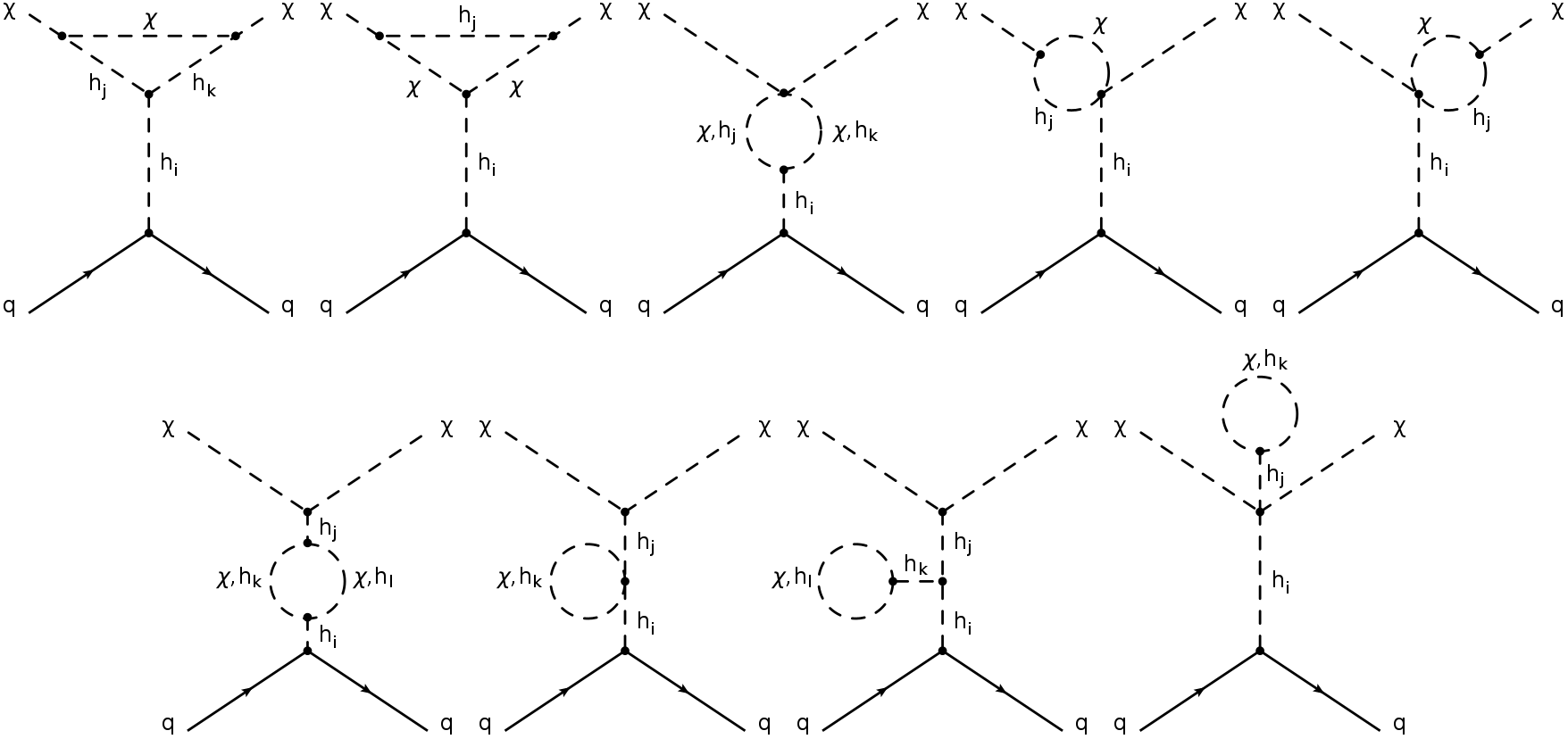}
\caption{One-loop topologies that contribute to
the DM-nucleon
scattering cross section in our approximation
with $q \in \{u,d,c,s,b,t\}$ and $i,j,k,l\in\{1,2,3\}$.}
\label{figtopos}
\end{figure}

With all the above considerations the set of diagrams that actually contribute to the one-loop cross-section 
is the one with the topologies depicted in \reffi{figtopos}, \htm{containing} the upper $h_i\chi\chi$-vertex and
the $h_i$-propagators \htm{corrections}.
As \htm{discussed} 
before, the only particles
that have to be considered
in the loops are
the neutral CP-even Higgs bosons
$h_{1,2,3}$ and the DM particle $\chi$,
since the diagrams with the other particles
cancel each other out as a result of the
U(1) symmetry.

\htm{Moreover, the above considerations
lead us to consider only 
the effective scalar operator for the computation of the scattering
cross sections of the DM on \htb{nucleons},
\begin{equation}
\mathcal{L_{\rm eff}} = m_q
C^s_q \chi \chi \bar q q \ ,
\label{eqeffop}
\end{equation}
where $m_q$ is mass of
the quark, and
$C^s_q$ is the Wilson coefficient
that is determined order by order
in perturbation theory from the
matching to the full model.}
Since one has to \htb{consider} the
scattering on both up-type quarks
and down-type quarks, there are important
differences between the different Yukawa
types of the S2HDM. In the type~I and the
type~LS (Lepton Specific), only the doublet field $\phi_2$
is coupled to the quarks, independently
of the quark flavour. As a result of the
fact that the dependence on the mass of
the different quarks is factored out of
the Wilson coefficients $C_q^s$ as shown
in \refeq{eqeffop}, in these types $C_q^s$ is
identical for all six quark flavours,
i.e.
\begin{equation}
C_q^{\rm I,LS} = C_{u,d,c,s,b,t}^s \ .
\label{eq:wilsons1}
\end{equation}
In contrast, in the Yukawa types~II
and~F (Flipped) the doublet field $\phi_2$ is
coupled to up-type quarks, and
the field $\phi_1$
is coupled to down-type quarks.
This gives rise to the fact that
the amplitudes are different depending
on whether the DM particle $\chi$ scatters
on up-type quarks or down-type quarks.\footnote{Also
the tree-level amplitudes \htm{given
in \refap{app:treeamp}} are different
in type~II and type~F depending on whether
$\chi$ scatters on up-type or down-type
quarks. However, at tree-level both
amplitudes vanish in the limit of
zero momentum transfer.}
Consequently, one finds two different
Wilson coefficients which we denote
\begin{equation}
C_u^{\rm II,F} = 
C_{u,c,t}^s (=
C_q^{\rm I,LS})
\quad \textrm{and}
\quad C_d^{\rm II,F} = C_{d,s,b}^s \ ,
\label{eq:wilsons2}
\end{equation}
in the following.

The calculations of the one-loop corrections
as described above were performed using {\tt FeynRules~2.3.48}~\cite{Christensen:2008py,Degrande:2011ua,Alloul:2013bka},
 {\tt FeynArts~3.10}~\cite{Kublbeck:1990xc,Hahn:2000kx} and {\tt FeynCalc~10.0.0}~\cite{Mertig:1990an,Shtabovenko:2016sxi}. 
 An independent calculation was performed using  {\tt SARAH~4.14.3}~\cite{Staub:2009bi,Staub:2010jh,Staub:2012pb,Staub:2013tta,Staub:2015kfa},
  {\tt FeynArts~3.11} and  {\tt FormCalc~9.9}~\cite{Hahn:2016ebn}
 All loop integrals were computed using LoopTools~\cite{Hahn:1998yk, vanOldenborgh:1989wn}.  We found agreement between
 both results.
\htm{As a consequence of the fact
that the total number of diagrams
is large, we refrain from giving analytic
expressions for the Wilson coefficients
$C_q^{\mathrm{I,LS}}$ and
$C_q^{\mathrm{II,F}}$ here, but instead
discuss their numerical impact in terms
of the DM-nucleon scattering cross sections,
as discussed in the following.
However, we make the obtained expressions
for the Wilson coefficients available to
the public as \texttt{Fortran} and
\texttt{python} routines.\footnote{The
routines are available at
\url{https://gitlab.com/thomas.biekoetter/dds2hdm}.
The computation of the DM-nucleon
scattering cross sections will
is also be implemented in the new version
of the public code
\texttt{s2hdmTools}~\cite{Biekotter:2021ovi}.}}

\subsection{From amplitudes to cross sections}
\label{sec:amp2xs}

The cross section  for  the scattering of $\chi$ on  \htb{nucleons}
as a function of $C_q^s$ can be expressed
as~\cite{Cerdeno:2010jj}
\begin{equation}
\sigma_{\chi N} = \frac{1}{\pi}
\frac{m_N^4}{(m_N + m_\chi)^2}
\left|
\sum_{q=u,d,s} C^s_q f^N_{Tq}
+ \frac{2}{27} f^N_{Tg}
    \sum_{q=b,c,t} C^s_q
\right|^2 \ ,
\label{eq:xsnucdm}
\end{equation}
with the DM mass $m_\chi$
and the nucleon mass $m_N$, and with
$N=n,p$ for neutrons or protons.
$C_q^s$ are the Wilson coefficients
of the effective DM-quark scattering
operators as defined in \refeq{eqeffop}.
\htm{The factors $f_{Tq}^N$ are
defined by the nucleon matrix elements}
as~\cite{Hisano:2012wm,Young:2009zb,Shifman:1978zn}
\begin{equation}
        \langle N| m_q\bar{q}q |N\rangle \equiv m_N f^N_{Tq} \,, 
\end{equation}
\htm{representing} the contributions
of the quarks to the
nucleon mass. Their numerical values
have been extracted from lattice
simulations and from data-driven
methods to be~\cite{Hisano:2015rsa,
Young:2009zb,Ohki:2012jyg,Owens:2012bv}
\begin{align}
f^p_{Tu} &= 0.029 \ , \;
f^p_{Td} = 0.027 \ , \;
f^p_{Ts} = 0.009 \ , \\
f^n_{Tu} &= 0.013 \ , \;
f^n_{Td} = 0.040 \ , \;
f^n_{Ts} = 0.009 \ .
\end{align}
The heavy quark contributions in \refeq{eq:xsnucdm}
\htm{are determined by making use
of the QCD trace anomaly}
that relates the heavy quark $Q=b,c,t$ operators with the 
gluon field strength tensor~\cite{Shifman:1978zn}
\begin{equation}
    m_Q \bar{Q}Q \rightarrow -\frac{\alpha_s}{12 \pi}  G_{\mu \nu} G^{\mu \nu}  \,.
    \label{eq::Mapping}
\end{equation}
The quantity $f_{Tg}^N$ is then defined
by the matrix element
\begin{equation}
         \langle N| -\frac{\alpha_s}{12\pi} G_{\mu \nu} G^{\mu \nu}  |N\rangle  \equiv \frac{2}{27} m_N f_{Tg}^N .
\end{equation}
$f_{Tg}^N$ can be expressed in terms
of the contributions of the
light quarks, such that~\cite{Cerdeno:2010jj}
\begin{equation}
f^N_{Tg} = 1 -
    \sum_{q=u,d,s} f^N_{Tq} \ .
\end{equation}
The first sum in \refeq{eq:xsnucdm}
\htm{running over the light quark
flavours $q=u,d,s$}
contains the contributions
from the scattering of $\chi$ directly
on the light quarks. 
Here the contributions from the heavy
quark flavours can be neglected because
their contributions to the nucleon mass are tiny
at the energy scales relevant for
the DM scattering on nuclei.
\htm{As described above,} the heavy quark contributions will be included as gluon initiated processes
making use of the QCD trace anomaly.
\htm{This gives rise
to} the second sum in \refeq{eq:xsnucdm}, \htm{which}
contains the contributions from the
scattering on the gluons, where we take
into account at leading order only
the quark-mediated contributions.
As a consequence of this approximation,
this contribution can also be expressed in terms of the
effective operator shown in \refeq{eqeffop}.
Here the important
contributions arise from the heavy quarks
whose couplings to the Higgs bosons are not
suppressed by tiny Yukawa couplings.
Thus, the second sum runs only
over the heavy quark flavours $b$, $c$ and $t$.

As previously discussed, 
in our computation of $C^s_q$
we only include the numerically
dominant corrections to the upper
vertex $h_i \chi \chi$ and the
$h_i$-propagator corrections, according
to the strategy also applied in
\citere{Azevedo:2018exj} for the pNG DM
model with a single Higgs doublet.
In this approximation,
in type~II and
type~F the amplitudes $C^s_q$
are different for the
up-type quarks $q=u,c,t$ and the
down-type quarks $q=d,s,b$, whereas
in type~I and type~LS 
they are independent of the
quark flavour. In the latter case,
one can simplify \refeq{eq:xsnucdm}
and write it as
\begin{equation}
\sigma_N = \frac{1}{\pi}
\frac{m_N^4}{(m_N + m_\chi)^2}
\left|
C^s_q
\right|^2 f_N^2 \ , \; \text{with} \;
f_N =
\sum_{q=u,d,s} f^N_{Tq}
+ 3\frac{2}{27} f^N_{Tg}
= 0.27 \ ,
\end{equation}
where $f_N$ is the nucleon form factor
that was used in \citere{Azevedo:2018exj}.

\section{Numerical impact in light of
current and future experiments}
\label{sec:num}
In this section we will present the
numerical analysis of the DM direct-detection
cross sections at the approximate one-loop level.
We will start our discussion
in \refse{sec:numgen} by analyzing whether
our expressions for the one-loop contributions
fulfil some theoretical requirements
\htm{that can be derived from symmetry
arguments} in order
to cross check our results.
In the second step, we present the results of
two parameter scan projections in the type~I and the type~II
of the S2HDM with the goal of determining whether
the DM scattering cross sections are sufficiently
enhanced at the loop level such that the presence
of the DM state $\chi$ could be
\htm{probed} 
at DM direct-detection experiments.

\subsection{General considerations}
\label{sec:numgen}
Due to the large number of diagrams that
give rise to finite contributions to the
DM-nuclei scattering cross sections in
the limit of zero-momentum transfer,
as discussed in \refse{sec:calcu},
the complete expressions for the loop
corrections are rather lengthy and
complicated, such that they can only be
evaluated numerically.
Nevertheless, the expressions have to fulfil
some basic requirements that can be derived
by means of symmetry arguments
(see \citere{Azevedo:2018exj}
for a discussion in the
pNG DM model with one Higgs doublet).
We will discuss here if these requirements
are met by our result. This will also provide
us with a first insight about the order of
magnitudes of the cross sections that can
be achieved in the S2HDM beyond tree-level.
A more complete assessment of the phenomenological
impact can be found in \refse{sec:numscan}, where we
will discuss two parameter scan projections in which
we take into account the whole list of
theoretical and experimental constraints
mentioned in \refse{sec:thecon} and
\refse{sec:expcon}, respectively.

The presence of non-vanishing corrections to
the scattering cross sections at the loop level
is related to the fact that the U(1) symmetry,
under which the singlet field $\phi_S$ is charged,
is softly broken in order to give rise to
a mass for the DM state $\chi$. If the U(1) symmetry
would be exact, the cancellation mechanism for
the $t$-channel Higgs-boson exchange between
$\chi$ and the quarks would hold at all orders
in perturbation theory.
A condition that the one-loop corrections have
to fulfil is therefore that in the limit of
$m_\chi \to 0$, i.e.~in the limit in which
the U(1) symmetry is restored, the corrections
have to vanish as well. On the other hand,
if the DM mass becomes much larger than the
masses of the Higgs bosons, i.e.~$m_\chi \gg m_{h_i}$,
the cross sections become smaller as
a result of the factor $1 / m_\chi^2$
in \refeq{eq:xsnucdm}.

\begin{figure}
\floatbox[{
  \capbeside
  \thisfloatsetup{
    capbesideposition={left,top},
    capbesidewidth=0.3\textwidth}}]{figure}[\FBwidth]
{\caption{\footnotesize
  DM-proton scattering cross section $\sigma_{\chi p}$
  in dependence of the DM mass $m_\chi$ for different
  values of the singlet vev $v_S$ in type~II.
  The other parameters
  are fixed to the values shown on the right.
  \htm{Also shown are the current upper limits on the
  $95\%$ confidence level from
  XENON1T~\cite{XENON:2018voc}
  (blue dashed),
  PandasX-4T~\cite{PandaX-4T:2021bab}
  (red dashed) and
  LZ~\cite{LZnew} (green dashed),
  and the projected upper limit from
  Darwin~\cite{DARWIN:2016hyl}
  (dotted).
  The gray area indicates the neutrino
  floor~\cite{Billard:2013qya}.}}
\label{fig:checkmDMzero}}
{
\includegraphics[width=0.68\textwidth]{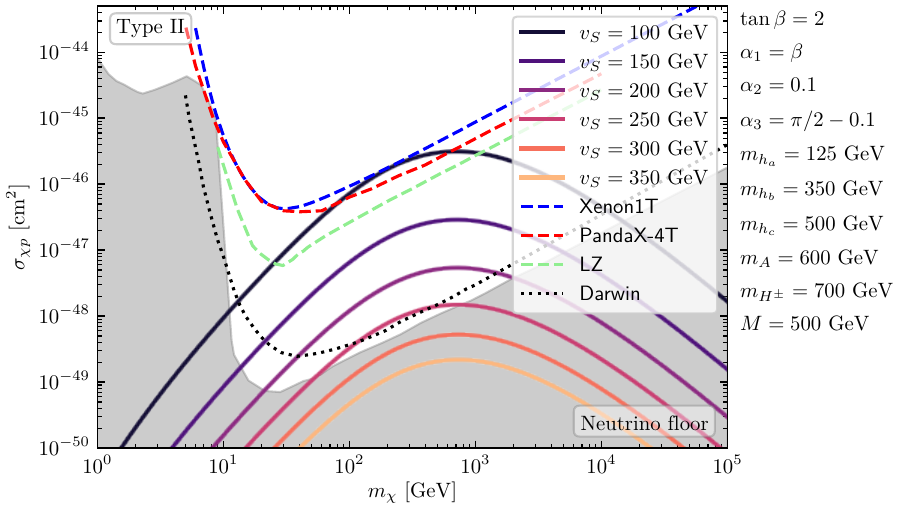}
}
\end{figure}

In \reffi{fig:checkmDMzero} we show the
predictions for the cross sections of the
scattering of $\chi$ on protons $\sigma_{\chi p}$
as a function of $m_\chi$ in the
type~II S2HDM. We show $\sigma_{\chi p}$
for different values of the singlet vev $v_S$,
where the value of the latter is indicated by
the color coding of the lines.
The values of the remaining free parameters
are given next to the plot on the right-hand side.
The parameter values were chosen such that
the theoretical constraints discussed in
\refse{sec:thecon}, in particular the perturbative-unitarity
constraints, are respected. However, we did not
apply the experimental constraints on the Higgs
sector and the DM sector.
\htm{Also shown with the dashed lines
are the exclusion limits at the
95\% confidence level from the XENON1T
experiment~\cite{XENON:2018voc} (blue),
the PandaX-4T experiment~\cite{PandaX-4T:2021bab}
(red) and the LZ
experiment~\cite{LZnew}, respectively,
and the dotted line indicates
the future projected
exclusion limits from the Darwin
experiment~\cite{DARWIN:2016hyl}.}
The gray shaded area indicates the neutrino
floor~\cite{Billard:2013qya}.
As expected based on the discussion above, the
cross sections vanish in the limit $m_\chi \to 0$
independently of the value of $v_S$.
$\sigma_{\chi p}$~reaches the maximum value
for DM masses that are close to the masses of
the CP-even Higgs bosons $h_i$. For DM masses
that are much larger the cross sections drop
again until they fall below the neutrino floor
at $m_\chi \gtrsim 10 \tev$ for the
smallest values of $v_S$ considered, whereas
for the largest values of $v_S$ the predictions
are always within the neutrino floor.
One can \htm{generically} observe
that overall larger values
of~$\sigma_{\chi p}$ can be achieved for smaller
values of $v_S$. This is due to the fact that
for fixed values of the masses $m_{h_i}$ smaller
values of $v_S$ give rise to larger values of
the quartic couplings $\lambda_{6,7,8}$. These couplings
act as the portal couplings between the
visible and the dark sector, such that larger values
of $\lambda_{6,7,8}$ give rise to larger values of
the scattering cross sections.
However, larger values of the quartic couplings also
yield larger values of the annihilation cross sections
and, therefore, smaller values of the predicted
relic abundance. As a consequence, the parameter points
with the largest values of $\sigma_{\chi p}$
\htm{can be expected to }predict
a relic abundance which is
smaller than the
measured DM relic abundance. The impact of the predicted
DM density on the prospects of probing the S2HDM
at \htm{DD} 
experiments will be discussed
in more detail in \refse{sec:numscan}.

By comparing the theoretical predictions with the
upper limit from XENON1T
\htm{PandaX-4T and LZ}, one can see that only
for the smallest value of $v_S = 100\gev$ considered here
\htm{the current DD experiments have }
the \htm{potential}
of probing
the S2HDM parameter space. It should be noted that
\htm{even} smaller values of $v_S$, for which $\sigma_{\chi p}$
would become even larger, are excluded in this scenario
as a consequence of the tree-level perturbative
unitarity constraints. This emphasizes the importance
of taking into account such theoretical constraints
in order to give an accurate estimate of the maximum
values of $\sigma_{\chi N}$ that can be achieved
in the S2HDM. While the current upper limits from
XENON1T \htm{PandaX-4T, and LZ}
barely constrain the parameter points shown
in \reffi{fig:checkmDMzero},
large parts of the interval of DM masses
that are shown can be probed in the future by Darwin.
For instance, assuming a value of $v_S = 200\gev$,
the expected limits from Darwin would exclude
the DM mass range $30\gev \lesssim m_\chi \lesssim 1.8\tev$.
In general it is interesting to note that
the scattering cross sections peak for DM masses
of the order of the masses of the Higgs boson.
The presence of the BSM Higgs bosons can be tested
at the LHC if they are not too heavy to be produced.
In this case the S2HDM can be probed in a complementary
way by \htm{DD} 
experiments and colliders.

\begin{figure}
\floatbox[{
  \capbeside
  \thisfloatsetup{
    capbesideposition={left,top},
    capbesidewidth=0.27\textwidth}}]{figure}[\FBwidth]
{\caption{\footnotesize
  DM-proton scattering cross section $\sigma_{\chi p}$
  in dependence of the mass of one of the
  CP-even Higgs bosons $m_{h_b}$ for different
  values of the singlet vev $v_S$ in type~I.
  The other parameters
  are fixed to the values shown on the right.}
\label{fig:checkmHszero}}
{
\includegraphics[width=0.67\textwidth]{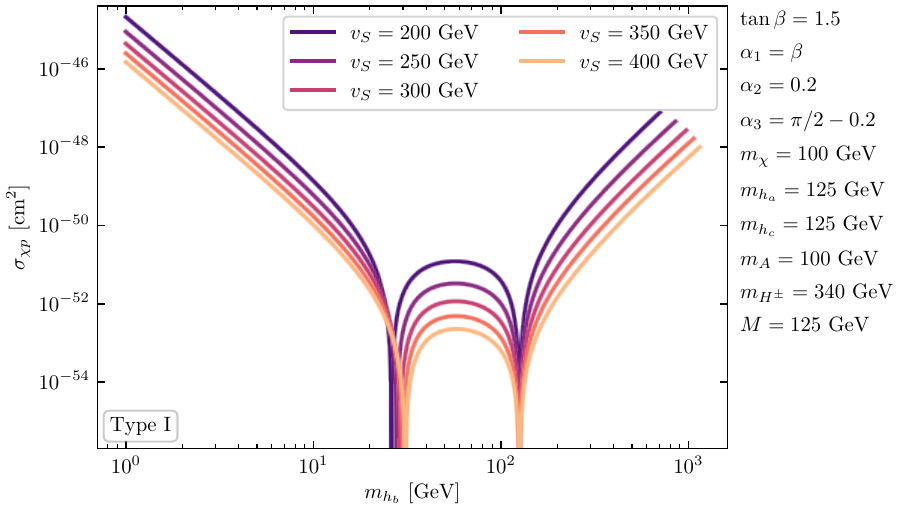}
}
\end{figure}

Another theoretical requirement which has to be
fulfilled by the one-loop corrections that we
take into account is that the cancellation mechanism
only holds in the limit of vanishing momentum
transfer. 
As a result, the cancellation mechanism breaks down if the mass of one of the Higgs bosons is not much larger than the momentum that is transferred in the scattering process.
In order to demonstrate that our result also
complies with this condition, we show in
\reffi{fig:checkmHszero} the predictions for
$\sigma_{\chi p}$ as a function of the mass
of one of the Higgs bosons $h_b$, with the
remaining parameters fixed to the values
shown on the right-hand side of the plot, and
where we show here the predictions of the
type~I S2HDM.
One can see that, as expected,
$\sigma_{\chi p}$ increases 
\htm{drastically}
in the limit $m_{h_b} \to 0$, independently
of the value of $v_S$ as indicated by the
color coding of the lines.
As before, we applied only the theoretical
constraints in order to produce the results
shown in \reffi{fig:checkmHszero}, whereas
the experimental constraints were not applied.
This is important to note because values of
$m_{h_b} \ll 125 / 2 \gev$ would be excluded
due to constraints from the signal-rate
measurements of $h_{125}$ in combination with
the condition of not overclosing the
universe~\cite{Biekotter:2021ovi}.
As a result, although the direct-detection
cross sections can be very large if the
singlet-like Higgs boson (here $h_b$)
is much lighter than
$125\gev$, DM direct-detection experiments
cannot provide additional exclusion limits
in this region of the parameter space.

In the opposite limit with $m_{h_b} \gg m_\chi$,
one can observe in \reffi{fig:checkmHszero}
that $\sigma_{\chi p}$
instead increases with increasing value of $m_{h_b}$.
This behaviour has its origin in the fact that
for fixed values of $v_S$ the
\htk{absolute values of the} quartic couplings
$\lambda_{6,7,8}$ grow with increasing value
of the singlet-like Higgs-boson mass $m_{h_b}$.
As already mentioned, larger absolute values
of the quartic couplings give then rise to
larger scattering cross sections.
The absolute values of the quartic couplings
are ultimately bounded from above by the
constraints from perturbative unitarity.
The predictions in \reffi{fig:checkmHszero}
are shown for each
value of $v_S$ up the maximum value of
$m_{h_b}$ for which the parameter points were
still in agreement with these bounds.
Consequently, the maximum values of $\sigma_{\chi p}$
that are achieved here in a parameter region
that is potentially not yet excluded by other
experimental constraints are of the order of
$\sigma_{\chi p} \sim 10^{-48} \ \mathrm{cm}^2$,
which is well within the range that can be tested
at future 
\htm{DD} experiments like Darwin.

Another interesting feature that can be observed
in \reffi{fig:checkmHszero} is the
appearance of \textit{blind-spots}
at certain values of $m_{h_b}$ where
$\sigma_{\chi p}$ drops to zero.
Such blind-spot regions were also observed in
the simpler case of \htm{the} pNG DM
\htm{model} with only one Higgs
doublet~\cite{Azevedo:2018exj}. 
\htm{The presence of the blind-spots is}
a result of a
cancellation between the amplitudes of different
loop diagrams, giving rise to the fact that the
sum of all amplitudes, and thus the Wilson
coefficients $C_q^s$, vanish.
For the blind-spot on the right-hand
side it is easy to see that it appears at
the point at which
all CP-even Higgs bosons are mass degenerate,
with $m_{h_{a,b,c}} = 125\gev$.
Even though it is questionable whether
such a situation is phenomenologically viable
in light of constraints from the LHC measurements,
it is still an interesting observation that
approximately mass-degenerate scalar states could
yield a highly suppressed DM\htm{-nucleon
scattering}
cross section.
A second blind-spot can be observed at roughly
$m_{h_b} \sim 30\gev$, where the precise location
depends on the value of $v_S$.
In addition, the location of this
additional blind-spot also
depends in a non-trivial way on the choice of
the masses $m_{h_{a,b,c}}$ and the mixing angles
$\alpha_{1,2,3}$. For both blind-spots,
it might be interesting
to compute corrections beyond the one-loop level
in order to analyze whether they would
remain, in which case their presence would be related
to an accidental symmetry, or whether the higher-order
corrections eliminate the blind-spots, in which case
their presence relies on a purely accidental choice
of parameters.

In addition to the blind-spots that appear
due to vanishing scattering amplitudes
between the DM state $\chi$ and the
quarks, as discussed above, in the
type~II and the type~F S2HDM further blind-spots
can appear as a result of a cancellation
between the different terms in the sum
over the quark contributions as shown
in \refeq{eq:xsnucdm}. As discussed in
\refse{secwilson}, in type~I
and type~LS (at the one-loop level) there
is only a single Wilson coefficient
$C_q^{\rm I,LS}$ that enters in this sum.
However, in type~II and type~F there
are two independent coefficients
$C_u^{\rm II,F}$ and $C_d^{\rm II,F}$
\htm{(see \refeq{eq:wilsons2} and the related
discussion)}
for the scattering on up-type
and down-type quarks, respectively.
If these two coefficients have the opposite
sign, the sum in \refeq{eq:xsnucdm}
can be suppressed
even though the individual terms are
unsuppressed.

\begin{figure}
\centering
\includegraphics[width=0.98\textwidth]{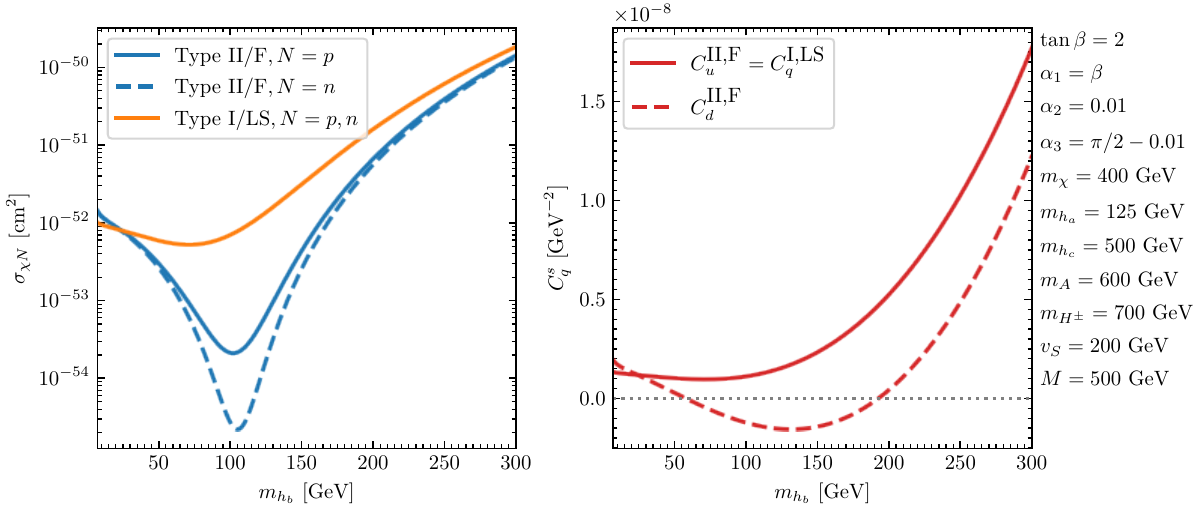}
\caption{\footnotesize
  Left: Cross sections for the scattering
  of $\chi$ on protons ($N=p$) and neutrons
  ($N=n$) as a function of $m_{h_b}$
  in type~I (orange) and type~II (blue). 
  Right: Wilson coefficients as defined
  in \refeq{eq:wilsons1} and
  \refeq{eq:wilsons2} as a function
  of $m_{h_b}$. The remaining parameters
  are fixed to the values shown on the right.}
\label{fig:ampliblinds}
\end{figure}

In order to demonstrate this
feature, we show in the left plot
of \reffi{fig:ampliblinds}
the cross sections for the scattering of
the DM state $\chi$ on protons and neutrons
in type~I (orange line) and type~II
(blue lines) for a representative benchmark
scenario. As before, we applied here only
the theoretical constraints in order to
ensure that the scalar potential is
well behaved. One can see that at
values of $m_{h_b} \sim 100\gev$ the
scattering cross sections in type~II
decrease by two orders of magnitude,
whereas the cross sections in the type~I
remains almost constant.
Moreover, it should be noted that
in this interval of $m_{h_b}$ the
cross sections in type~II are substantially
different for
the scattering on protons (solid blue line)
and neutrons (dashed blue line).
On the other hand, in type~I both
cross sections are practically equal, and consequently
only one line for both the scattering on
protons and on neutrons is shown.
As a phenomenological consequence, one
can notice that since different nuclei
are composed out of a different number of
neutrons and protons, a hypothetical
measurement of the scattering
cross sections
on different kinds of nuclei could be utilized to
distinguish between a DM candidate $\chi$
as predicted by the types~I/LS or
the types~II,F, respectively.

The suppression of the cross sections
in type~II can
be understood by the fact that one
of the Wilson coefficients
$C_u^{\rm II,F}$ or $C_d^{\rm II,F}$
changes the sign at the corresponding
mass interval of $h_b$.
In the right plot of \reffi{fig:ampliblinds}
we show the Wilson coefficients
as a function of $m_{h_b}$ for the
same benchmark scenario as was used
in the left plot of \reffi{fig:ampliblinds}.
As expected, one can see that
\htk{one of} the
coefficient\htk{s ($C_d^{\rm II,F}$,
dashed line)} becomes
negative in the 
\htm{mass range} $50\gev \lesssim
m_{h_b} \lesssim 200\gev$, where 
\htm{the mass range}
coincides with the \htm{one} 
in the left plot
in which the cross sections in type~II
are strongly suppressed.
Since in type~I there is only one Wilson
coefficient $C_q^{\rm I,LS}$, which is
identical to the coefficient $C_u^{\rm II,F}$
in type~II (solid line), the change of
the sign of $C_d^{\rm II,F}$ has no
impact on the cross sections in type~I.
Finally, we note that the precise
location of the blind-spot visible for type~II
and also the amount of the suppression
of the cross sections depend on the
nucleon form factors $f_{T_q}^N$, which are
only known approximately as they
are determined
from lattice simulations and from experimental
data. As a consequence, in the
parameter regions in which the scattering
cross sections are suppressed
due to the accidental cancellation
of contributions
from different quark types with opposite sign,
the 
\htm{relative} uncertainty of the
cross-section predictions associated to the
uncertainty of the form factors should be
regarded as larger compared to other
parameter space regions in which no such
cancellation takes place.

As a summary of the discussion in this section,
one can conclude that the one-loop corrections
included in our computation fulfil the
theoretical requirements that can be derived
from symmetry arguments, which serves as
a non-trivial cross-check of our results.
Moreover, we have demonstrated
that the cross sections as predicted at
the one-loop level can be well within the
reach of future DM direct-detection experiments.
It should be noted that we did not apply
here the experimental constraints on the model
parameters as introduced in \refse{sec:expcon}.
In order to verify whether the
future sensitivity of DM direct-detection
experiments is capable of probing parameter
space regions that are not yet excluded by
other experimental constraints on the Higgs
sector and the DM sector of the S2HDM, we
will discuss in the following section two
parameter scans in the type~I and the
type~II S2HDM in which the experimental
constraints will be taken into account.

\subsection{Parameter scans in type~I and type~II}
\label{sec:numscan}
In order to estimate the relevance of the
loop-corrected predictions for the cross
sections of the scattering of the DM state
$\chi$ on nuclei, we present here the predictions
in two parameter scan projections in the S2HDM type~I
and type~II in which we take into account
all the theoretical and experimental constraints
discussed in \refse{sec:thecon} and
\refse{sec:expcon}, respectively.
We note here that the Yukawa sectors
of type~I and type~LS as well as the Yukawa
sectors of type~II and type~F
only differ in the couplings of the
Higgs bosons to leptons.
Consequently, the cross-section predictions
for the DM-nucleon scattering
in the type~I are identical to the
predictions in the type~LS, and
the predictions in the type~II are identical
to the ones in the type~F.
Accordingly, apart from the different
collider constraints that have to be applied,
the results using type~I and~II
presented in the following also
provide a good understanding of the
importance of future DM \htm{DD} 
experiments in the type~LS and the type~F.

\begin{table}
{\renewcommand{\arraystretch}{1.4}
\footnotesize
\begin{flushleft}
\begin{tabular}{l|cccccccc}
Type & $m_{h_a}$ & $m_{h_b},m_{h_c},m_A,m_\chi$ &
  $m_{H^\pm}$ & $\alpha_{1,2,3}$ 
    & $\tan\beta$ & $M$ & $v_S$ \\
\hline
\hline
I & 125.09 & [30,1000] & [150,1000] &
  $[-\pi / 2, \pi 2]$ & [1.5,10] &
    [20, 1000] & [30,1000] \\
\end{tabular}
\newline
\vspace{0.8em}
\newline
\begin{tabular}{l|cccccccc}
Type & $m_{h_a}$ & $m_{h_b},m_A$ & $m_{H^\pm}$ &
  $m_{h_c,\chi}$ & $\alpha_{1,2,3}$ 
    & $\tan\beta$ & $M$ & $v_S$ \\
\hline
\hline
II & 125.09 & [200,1000] & [650,1000] & [30,1000] &
  $[-\pi / 2, \pi 2]$ & [1.5,10] &
    [450, 1000] & [30,1000] \\
\end{tabular}
\end{flushleft}
}
\caption{\footnotesize
  Values of the free parameters for the
  scan in type~I (top) and type~II (bottom).
  Dimensionful parameters are given in GeV.
}
\label{tab:vals}
\end{table}

In our scans we used values for the
free parameters as shown in \refta{tab:vals}.
We fixed $m_{h_a} = 125.09\gev$ in order to
account for a scalar state that could,
depending on its couplings,
\htm{behave in agreement with the
experimental measurements
with regards to} the discovered
Higgs boson. The masses of the BSM scalars
were scanned up to values of $1\tev$,
corresponding to a range that 
\htm{is potentially in reach of} the LHC.
It should be noted here that for the
scan in type~II we used
a lower limit of $m_{H^\pm} > 650\gev$ in
order to bypass constraints from flavour-physics
observables, whereas in type~I we used a lower
limit of $m_{H^\pm} > 150\gev$ since the flavour
constraints are much weaker (see also the
discussion in \refse{sec:expcon}).
\htm{In combination with} the theoretical
constraints on the quartic scalar couplings
and constraints from the EWPO, also
the lower limits on
the mass scale $M$ and the masses
$m_{h_b}$ and $m_A$ of
one of the CP-even scalars $h_b$ and the
pseudoscalar $A$, respectively,
\htm{are pushed to larger values in type~II}
in order
to account for the fact that the differences
between these parameters and $m_{H^\pm}$
cannot be too large.
The mixing angles were scanned over all
physically distinguishable parameter space,
\htm{and the lower limit on $\tan\beta$
was chosen according to constraints
from flavour physics.}
Finally, the singlet vev $v_S$
is varied within the scan range of the
BSM scalars.
\htm{We note that due to its pNG nature
the DM state $\chi$ can be light even
though the global U(1) symmetry breaking
has its origin at energy scales much larger
than the TeV scale, such that
also values of~$v_S \gg 1\tev$ would be
physically reasonable.}
However, \htm{as} we
\htm{demonstrated} in \refse{sec:numgen},
sizable values
of the cross sections for the
\htm{scattering of the} 
$\chi$ on nuclei are present only if
$v_S$ is of the order of the masses of
the CP-even Higgs bosons or smaller.
Therefore, for the purpose of determining
the largest scattering cross sections
that can be realized in the S2HDM it is
sufficient to scan only a range in which
$v_S$ is of the order of $m_{h_{a,b,c}}$
\htm{(or smaller)}.

We have generated parameter points by scanning
uniformly over the given parameter ranges.
For each parameter point generated in
this way, we have applied the theoretical and
experimental constraints discussed above,
and we have discarded the parameter
points for which one of the constraints
was violated.
For the remaining parameter points, we
have calculated the predictions for the
DM-nucleon scattering cross sections.
\htm{We have then} compared the theoretical
predictions against the current and
future DM direct-detection constraints
from the XENON1T, PandaX-4T, LZ
and the Darwin experiment,
respectively.
\htm{Finally}, we have also taken into
account the predicted value of the DM
density as obtained by assuming the
standard freeze-out mechanism in order
to answer the question whether the parameter
points that could be probed by 
\htm{DD} 
experiments would also
predict a sizeable fraction of the
measured DM relic abundance.
Moreover, in case the predicted relic
abundance is substantially smaller,
we address how much this reduces
the prospects of probing the
corresponding S2HDM parameter space
\htm{by means of DD} 
experiments.

\begin{figure}[t]
\centering
\includegraphics[width=0.48\textwidth]{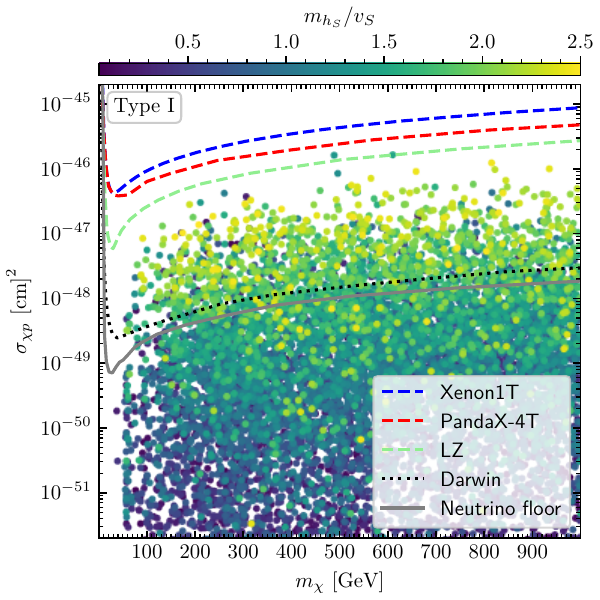}~
\includegraphics[width=0.48\textwidth]{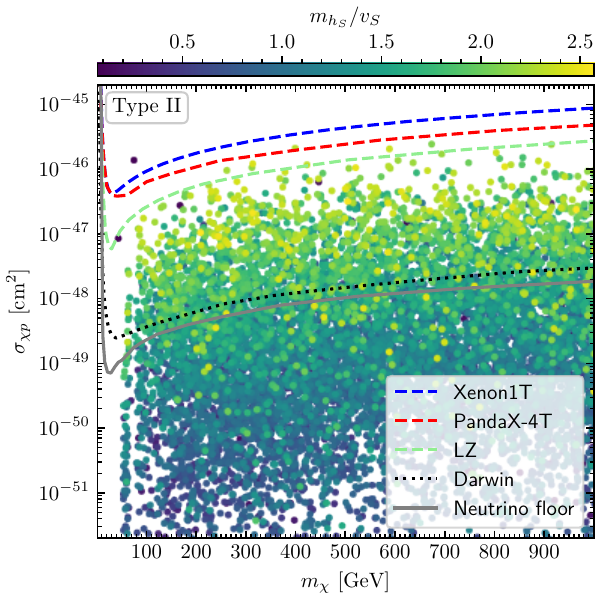}
\includegraphics[width=0.48\textwidth]{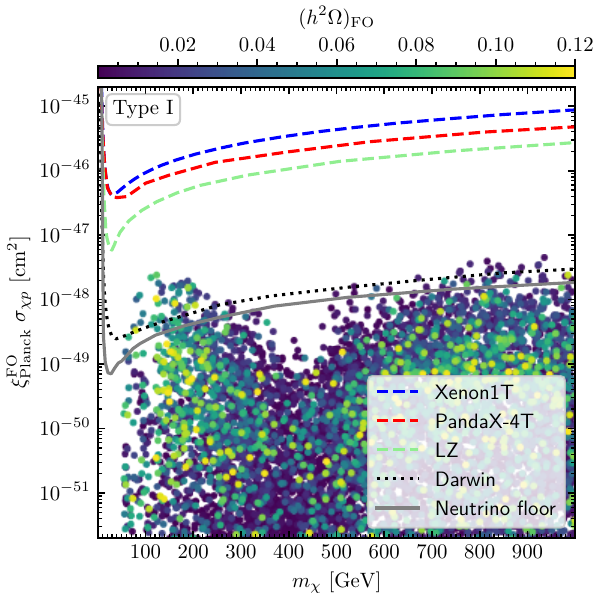}~
\includegraphics[width=0.48\textwidth]{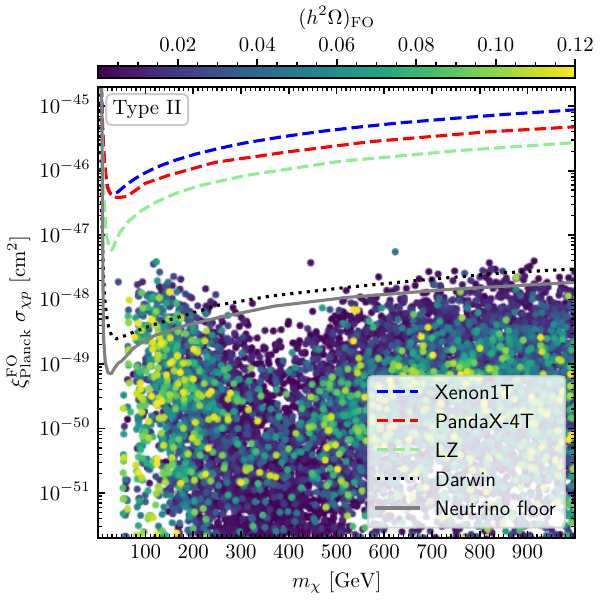}
\caption{\footnotesize
  Parameter points of the scan in type~I (left)
  and type~I (right) in the $(m_\chi$,$\sigma_{\chi p})$
  plane (top) and in the
  $(m_\chi$,$\xi_{\rm Planck}^{FO}\sigma_{\chi p})$
  plane (bottom). The color coding of the points
  indicates the value of $m_{h_S} / v_S$
  (top) and the value of $(h^2 \Omega)_{\rm FO}$
  (bottom). Also shown are the current upper limits
  on the $95\%$ confidence level from
  XENON1T~\cite{XENON:2018voc}
  \htm{(blue dashed line), from
  PandaX-4T~\cite{PandaX-4T:2021bab}
  (red dashed line) and from
  LZ~\cite{LZnew} (dashed green line),}
  and the projected upper
  limit from
  Darwin~\cite{DARWIN:2016hyl}
  (dotted line).
  The gray solid line indicates the
  neutrino floor~\cite{Billard:2013qya}.}
\label{fig:scan}
\end{figure}

In the top row of
\reffi{fig:scan} we show the scan points
in type~I (left) and type~II (right) with
the DM mass $m_\chi$ on the horizontal axis
and the DM-proton scattering cross section
$\sigma_{\chi p}$ on the vertical axis.
The color coding of the points indicates
the value of $m_{h_S} / v_S$, where $h_S$
is defined as the CP-even scalar $h_i$ with
the largest singlet admixture
\htm{given by}~$R_{i3}^2$
(see \refse{sec:modeldefs}).
Also indicated are the cross section limits
at the 95\% confidence level from the
XENON1T~\cite{XENON:2018voc}, \htb{the PandaX-4T~\cite{PandaX-4T:2021bab}
and the LZ~\cite{LZnew} experiments} 
with \htb{blue, red and green dashed lines, respectively}, and the projected
future limits from the Darwin
experiment~\cite{DARWIN:2016hyl} with
the black dashed line.
Finally, the gray solid line indicates
the neutrino floor~\cite{Billard:2013qya}.
One can see that we find points which predict
values of $\sigma_{\chi p}$ that are within
the reach of Darwin, whereas the current
experimental sensitivity by XENON1T,
\htm{PandaX-4T and LZ are} 
not sufficient
to probe the S2HDM parameter space
\htm{in a significant way}.
On the other hand, the largest fractions
of parameter points feature values of
$\sigma_{\chi p}$ \htb{that are} substantially below
the Darwin sensitivity, and many points
are within the neutrino floor in which case
a possible DM detection is not very promising
even in the distant future.
We note here that the range of the
vertical axis for $\sigma_{\chi p}$ was
set to $10^{-52} \ \mathrm{cm}^2$ for
a better visibility of the relevant range
\htm{of~$\sigma_{\chi p}$}
for which there is experimental sensitivity,
although there are parameter points featuring
values of $\sigma_{\chi p}$ that are orders
of magnitude smaller.
Finally, we emphasize that overall larger
values of the DM-proton scattering cross section
are correlated with larger values of
the ratio $m_{h_S} / v_S$, which is in
agreement with the observations discussed
in \refse{sec:numgen}.

The cross-section limits from XENON1T,
\htb{PandaX-4T, LZ} and
Darwin, as shown in \htb{the top row of} \reffi{fig:scan},
were derived under the assumption
that the DM particle under consideration
accounts for the entire measured DM relic
density as measured by Planck.
However, in the S2HDM one can
predict the DM relic abundance
composed of the state $\chi$ assuming the
standard freeze-out scenario, and in
the parameter scans
we only demanded that the 
predicted DM relic abundance is not
larger than the measured value, thus leaving
room for additional sources that contribute
to the DM relic abundance.
If the predicted abundance of
the DM state $\chi$ is smaller
than the measured value, the prospects
for the \htm{DD} 
of DM decrease, since
the number of scattering events in the
detector is smaller compared to the
number of scattering events expected
based on the measured DM density.
In order to account for the impact
of the predicted relic abundance, it is
illustrative to
compare the upper limits on the scattering
cross section from 
\htm{DD} 
experiments against
the predicted scattering cross section
$\sigma_{\chi p}$ times a scaling factor
\begin{equation}
\xi_{\rm Planck}^{\rm FO} = \frac
{\left(h^2 \Omega \right)_{\rm FO}}
{\left(h^2 \Omega \right)_{\rm Planck}} \ ,
\end{equation}
where $(h^2 \Omega)_{\rm FO}$ is the
theoretical prediction for 
today's
DM relic abundance based on the freeze-out
mechanism (obtained with the help of
\texttt{MicrOmegas}), and
$(h^2 \Omega)_{\rm Planck} = (0.119
\pm 0.003)$ is the value as measured
by \htm{the} Planck
\htm{satellite}~\cite{Planck:2018vyg}.

In the bottom row of \reffi{fig:scan} we show
the rescaled cross sections
$\xi_{\rm Planck}^{\rm FO} \sigma_{\chi p}$
in dependence of the DM mass $m_\chi$
for type~I on the left and for type~II
on the right\htm{, respectively}.
Here the color coding indicates
the value of \htm{the predicted
DM relic abundance~}$(h^2 \Omega)_{\rm FO}$.
Since we demanded
$(h^2 \Omega)_{\rm FO} \leq
(h^2 \Omega)_{\rm Planck}$
\htm{(see \refse{sec:expcon})}, the parameter
points all feature $\xi_{\rm Planck}^{\rm FO}
\leq 1$. 
\htm{Thus,} compared to the plots
in the upper row of \reffi{fig:scan}, the points
move towards 
\htm{the neutrino floor and away from
the experimental upper limits
on the DM-nucleon scattering cross section.}
Nevertheless, we find a mass interval
$60 \gev \lesssim m_\chi \lesssim 300\gev$
in which Darwin has the potential to
probe the S2HDM parameter space
\htm{in both type~I and type~II}.
One should note that many of
the parameter points in
this interval of $m_\chi$ predict a sizable
fraction (or all) of the measured DM
relic abundance. Hence, the 
\htm{DD}
constraints will have the potential to
probe regions of the parameter space
that are especially interesting in view
of the predictions for
$(h^2 \Omega)_{\rm FO}$.\footnote{DM masses
of $63\gev\lesssim m_\chi \lesssim 67\gev$
were also shown
to \htb{be} favoured for a simultaneous
description of the
Fermi-LAT galactic-center excess and the
AMS antiproton excess in the
S2HDM~\cite{Biekotter:2021ovi}.}
For larger DM masses, additional \htm{DM} annihilation
channels, for instance into pairs of
on-shell vector bosons, top quarks or
Higgs bosons $h_i$, become kinematically
open. As a consequence, in the range
$\htk{300}\gev
\lesssim m_\chi \lesssim 500\gev$ we find a
strong suppression of $(h^2 \Omega)_{\rm FO}$,
and therefore $\xi_{\rm Planck}^{\rm FO} \ll 1$.
This gives rise to the fact that in this
range of $m_\chi$ almost no points are found
above the \htm{projected} upper limit of Darwin.
For values of $m_\chi \gtrsim 500\gev$,
one can see that
parameter points featuring sizable values
of $(h^2 \Omega)_{\rm FO}$ can be found
above the neutrino floor, however also
here the projected sensitivity of Darwin
is small and limited to parameter points
for which the DM state $\chi$ does not account
for the whole DM relic abundance.
Finally, we note that no large differences
between both Yukawa types can be found.
Accordingly, the prospects for probing
the S2HDM parameter space at future
DM direct-detection experiments can be
expected to be fairly similar.

\section{Conclusions}
\label{conclu}

In this work we have calculated the one-loop electroweak corrections to the DM-nucleon scattering cross section in the 
extension of the SM with one extra \htk{scalar}
doublet and one extra complex \htk{scalar}
singlet, dubbed S2HDM. The model provides a DM
candidate through a $U(1)$ 
\htm{symmetry} softly broken by dimension two terms. The tree-level amplitude of the DM-nucleon
process is proportional to the DM velocity
\htm{and thus negligible for direct-detection
experiments}. This is a feature of a class of models with a pseudo Nambu-Goldstone boson
as the DM candidate. 

The calculation was \htk{carried out} by two independent calculations
\htk{and exact agreement was found}.
Furthermore, from the theoretical point of view, 
the Nambu-Goldstone nature of the DM particle would have to be reflected on a zero cross section in the limit where 
the exact $U(1)$ symmetry is recovered. In fact, by making use of the non-linear formulation, it becomes clear
that for low-energy processes, the Nambu-Goldstone nature of the DM particle \htm{is} 
manifest due to
the proportionality between its mass and the $U(1)$ soft breaking parameter.
\htk{We explicitly checked that our results
vanish in the limit of zero DM mass and hence
comply with the above mentioned symmetry arguments.
We furthermore verified}
that there was no need to introduce
counterterms as the process is zero at tree-level
in the limit of zero
DM velocity. 
\htk{The fact that our results agree
with the features mentioned above} was an excellent
and non-trivial
cross-check of our computations.

A scan of the model parameters has been performed taking into account all theoretical constraints
\htm{that ensure that the electroweak vacuum is stable
and that the perturbative treatment of the
parameter points is valid. We have also
taken into account} the restrictions on the parameter space imposed
by the most relevant experimental constraints
\htm{related to the Higgs sector and the
DM sector of the S2HDM}. No parameter
points have been found that could be probed by
present direct detection
experiments \htm{such as} XENON1T, PandaX-4T
\htm{or LZ, while at the same time predicting a sizable fraction
of the measured DM relic abundance.}
However, \htm{we have found such parameter}
points within the reach of future experiments such as Darwin.
\htm{For these parameter points,
we have demonstrated that
they are characterized
by overall larger
values of the ratio $m_{h_S} / v_S$, where
$m_{h_S}$ is the mass of the Higgs boson with
the largest singlet admixture, and
$v_S$ is the U(1)-breaking vacuum expectation value
of the radial component of the singlet field.}
Hence, although the direct detection 
\htm{relies on a purely loop-induced} process \htm{in the S2HDM}, 
the cross sections \htk{can be} 
far from being too small to be probed in the future, \htm{in particular
if there is no larger hierarchy between
the electroweak scale and the energy scale
at which the breaking of the global
U(1) symmetry has its origin.}

\htm{In our parameter scans in type~I
and type~II we have not observed any major
differences regarding the parameter space
that could be probed in future direct-detection
experiments. Moreover, no major difference between
the maximum values of the DM-nucleon scattering
cross sections that can be achieved have
been found. However,} contrary to the
complex singlet extensions of the SM
as well as vector DM models,
the S2HDM has a very interesting
feature - the matrix elements
responsible for the scattering
cross section may depend
on the nucleon type. This is a consequence
of the existence of four different types of
Yukawa \htm{interactions, and that in the
type~I and~LS both up- and down-type quarks
are coupled to the same Higgs doublet,
whereas in type~II and~F the down-type
quarks are coupled to $\phi_1$ and the
up-type quarks to $\phi_2$, respectively.
This feature would potentially make it possible
to distinguish between the different
Yukawa types of the S2HDM if in the future the
scattering of DM will have been measured
on different kind of nuclei.}

\htm{Finally, we note that}
there are many other extensions with pNG bosons as dark matter candidates. The results obtained for this model show that most probably 
other extension with more fields will have
more freedom and may lead to
\htk{even} larger cross sections
\htm{once radiative corrections are considered}.

\section*{Acknowledgements}
\htm{We thank Sven Heinemeyer for the
invitation to participate in the HiggsDays21
workshop, during which the idea for
this project has been developed.}
We thank Mathias Pierre for interesting discussions.
The work of T.B.~is supported by the Deutsche
Forschungsgemeinschaft under Germany’s Excellence
Strategy EXC2121 ``Quantum Universe'' - 390833306.
The work T.B.~and M.O.O.~has been partially
funded by the Deutsche Forschungsgemeinschaft 
(DFG, German Research Foundation) - 491245950.
P.G. and R.S. are supported by FCT under contracts UIDB/00618/2020, UIDP/00618/2020, PTDC/FIS-PAR/31000/2017, 
CERN/FISPAR/0002/2017, CERN/FIS-PAR/0014/2019. 

\appendix
 
\htm{
\section{DM-quark scattering amplitude at tree-level}
\label{app:treeamp}
At the tree level, the scattering of $\chi$
on a quark $q$ is transmitted via the $t$-channel
exchange of the Higgs bosons $h_i$.
The corresponding amplitude $\mathcal{M}$
can be written as
\begin{equation}
\mathcal{M} =
- \frac{Y_q}{\sqrt{2}}
\sum_{i=1}^3
\frac{R_{ia} \ii \Gamma_{h_i\chi\chi}}
{m_{h_i}^2 + t} \ ,
\end{equation}
where $Y_q$ is the Yukawa coupling of the
quark $q=\{u,d,c,s,t,b\}$, $\Gamma_{h_i\chi\chi}$
is the tree-level coupling between
the DM particle $\chi$ and the Higgs bosons
$h_i$ given by
\begin{equation}
\ii \Gamma_{h_i\chi\chi}=
\lambda_7 v_1 R_{i1} +
\lambda_8 v_2 R_{i2} +
\lambda_6 v_S R_{i3} \ ,
\end{equation}
and $R_{ia}$ are the elements of the
mixing matrix of the CP-even scalars
defined in \refeq{mixingmatrix}, with $a=1$ or $a=2$
depending on whether the quark $q$ is coupled
to the doublet field $\phi_1$ or $\phi_2$, respectively.
Hence, in type~I and type~LS $a=2$ for
$q=\{u,d,c,s,t,b\}$, whereas in type~II and type~F
$a=2$ for $q=\{u,c,t\}$ and $a=1$ for $q=\{d,s,b\}$.
Rewriting the amplitude $\mathcal{M}$ in terms
of the squared masses, or vice-versa replacing the
squared masses in terms of the Lagrangian parameters
and the vevs, and by making use of the orthogonality
of $R$, it is easy to show that $\mathcal{M}$ vanishes
in the limit of zero-momentum exchange, i.e.~$t \to 0$.
}

\bibliographystyle{JHEP}
\bibliography{lit}

\newpage

\end{document}